\documentclass[twocolumn,preprintnumbers,nofootinbib,prl]{revtex4}
\usepackage{graphicx}

\usepackage[hypertex]{hyperref}
\usepackage{amsmath, amssymb}
\newcommand{\vev}[1]{\langle {#1} \rangle}
\newcommand{\lsim}{\lesssim}
\newcommand{\ord}[1]{\mathcal{O}{(#1)}}

\newcommand{\gsim}{\gtrsim}
\newcommand{\zp}{Z^\prime}
\newcommand{\krp}{k r_c \pi}
\newcommand{\beq}{\begin{equation}}
\newcommand{\eeq}{\end{equation}}
\newcommand{\fb}[1]{{#1}~fb$^{-1}$}

\begin{document}

% page numbers bottom-center
\pagestyle{plain}

\title{Big Signals of Little Randall-Sundrum Models}

\author{Hooman Davoudiasl\footnote{email: hooman@bnl.gov}}

\author{Shrihari Gopalakrishna\footnote{email:shri@quark.phy.bnl.gov}}

\author{Amarjit Soni\footnote{email: soni@bnl.gov}}

\affiliation{Department of Physics, Brookhaven National Laboratory,
Upton, NY 11973, USA}

%%%%%%%%%%%%%%%%%%%%%%%%%%%%%%%%%%%%%%%%%%%%%%%%%%%%%%%%%%%%%%%%%%%%%%%%%%%%

\begin{abstract}

We examine signals at the Large Hadron Collider (LHC) 
of Kaluza-Klein modes, in
volume-truncated ``Little Randall-Sundrum" (LRS) models of flavor,
characterized by 5D cutoff scales $M_5$ that are small compared to the 4D
Planck mass $M_P \sim 10^{19}$~GeV.  In particular, for the
phenomenologically viable choice $M_5 \sim 10^4$~TeV, the discovery
of a 2 (3)-TeV ``Little" $\zp$ at the LHC requires about 1 (4) fb$^{-1}$ at
$\sqrt{s}=10\, (14)$~TeV, {\it in the clean di-lepton channel}.  Our
results highlight the possibility of probing interesting values of
$M_5$, starting with the early LHC data.  With $M_5 \sim 10^4$~TeV,
discovering the second KK mode $Z^{\prime\prime}$, at about 4.3~TeV,
requires $\ord{100}$ fb$^{-1}$ at $\sqrt{s}=14$~TeV, providing a
probe of the warped nature of the bulk that is encoded in the mass
ratio of the first two KK modes, at design luminosity. By
comparison, discovering a 3-TeV $\zp$ of the Planck-weak hierarchy
models (with $M_5 \sim M_P$), {\it in any channel}, would require
upwards of $\ord{300}$ fb$^{-1}$ at $\sqrt{s}=14$~TeV. We also point
out that discovery prospects markedly improve for Little KK
gluons as well, but the challenging reconstruction of their $t {\bar t}$ 
decay products may not allow an appreciable advantage for early
discovery, over the Little $\zp$ case.

\end{abstract}
\maketitle

%%%%%%%%%%%%%%%%%%%%%%%%%%%%%%%%%%%%%%%%%%%%%%%%%%%%%%%%%%%%%%%%%%%%%%%%%%%
\underline{\it Introduction}: The Higgs condensate, $\vev{H} \simeq 250$~GeV,
sets the electroweak scale in the Standard Model (SM).
Yet, the SM Higgs potential is unstable against quantum corrections from the cutoff scale,
often assumed to be near the Planck scale $M_P \sim 10^{19}$~GeV.
This gives rise to the hierarchy problem, which is the puzzling smallness of the
ratio $\vev{H}/M_P \sim 10^{-17}$.  The Randall-Sundrum (RS) model
\cite{Randall:1999ee} was initially proposed to explain the hierarchy
by gravitationally red-shifting the 5D fundamental scale $M_5 \sim M_P$ down
to the TeV-scale, along a warped 5$^{\rm th}$ dimension.  This geometry is based on
a slice of AdS$_5$, truncated by flat 4D boundaries often referred to as the UV (Planck)
and IR (TeV) branes.  The RS metric is given by \cite{Randall:1999ee}
\beq
ds^2 = e^{-2 \sigma} \eta_{\mu\nu} dx^\mu dx^\nu - r_c^2 d\phi^2,
\label{metric}
\eeq
where $\sigma = k r_c |\phi|$, $k\lsim M_5$ is the 5D curvature scale,
$r_c$ is the radius of compactification, $-\pi\leq\phi\leq\pi$,
and a $\mathbb Z_2$ orbifolding of the 5$^{\rm th}$ dimension is assumed.  Henceforth,
we will assume $\phi \in [0, \pi]$ in our calculations.

In the original model, it was assumed that all
SM fields, and in particular the Higgs, are localized on the IR brane,
where the 5D parameter $\vev{H}_5 \sim M_5$ is exponentially red-shifted:
$\vev{H} = e^{-kr_c \pi} \vev{H}_5$.  For $kr_c \approx 12$,
the Planck-weak hierarchy is then generated .
The distinct signature of this model is
weak scale spin-2 Kaluza-Klein (KK) excitations
of the graviton, appearing as resonances in high energy
collisions \cite{Davoudiasl:1999jd}.  However, the cutoff scale in the 4D
effective theory is also red-shifted to near the weak-scale, leading to
unsuppressed higher dimensional operators, such as those for
unwanted flavor-changing neutral currents.

The situation can be improved by noting that the resolution of the
hierarchy only requires the Higgs to be
localized near the IR brane \cite{Goldberger:1999wh}
and SM gauge \cite{Davoudiasl:1999tf,Pomarol:1999ad}
and fermion \cite{Grossman:1999ra}
fields could propagate in the 5D bulk.
A mild modulation of bulk fermion masses provides
a natural mechanism for generation of SM fermion masses
and also suppression of unwanted 4-fermion operators \cite{Gherghetta:2000qt}.  This
is a result of the exponential localization of
fermion zero modes.  Small 4D Yukawa couplings
are naturally obtained, if light flavor zero modes are UV brane localized, and
operators containing light flavors are suppressed by scales much larger than $\vev{H}$.

While it is quite desirable to have a simultaneous resolution of
hierarchy and flavor puzzles,
the experimental signals of these warped models are now much more challenging.  For example,
localizing the light fermions near the UV brane
suppresses their couplings to IR-localized KK modes.  Therefore, KK production via light fermions
and decay into light clean di-lepton final states are suppressed.
This is a generic feature of warped models of hierarchy and flavor.

Even though localization of fermions alleviates many of the constraints on warped models,
precision data still require additional symmetries \cite{custodial} in order for
the new KK states to be as light as 2-3~TeV \cite{Carena:2007ua}.
Generally speaking, it has been shown that
a very likely new state in these models to be discovered at the LHC
is the first KK gluon \cite{KKgluon1,KKgluon2}.  The analysis
of Refs.~\cite{KKgluon1,KKgluon2} suggests that KK gluons as
heavy as 4~TeV will
be within the reach of the LHC.  However,
for the KK modes of the weak sector, the corresponding
reach is in the 2-3~TeV range \cite{Zprime,Agashe:2008jb} and typically requires several hundred
fb$^{-1}$.  For gauge KK masses in the above ranges, the graviton KK modes are most likely not
accessible, even with an upgraded LHC luminosity \cite{KKgraviton1,KKgraviton2}.
Of course, even the gauge KK sector remains barely accessible and would, in any event,
require a very large integrated luminosity~\cite{LHC-KK}.

In this work, we will 
concentrate on examining the experimental prospects for the
discovery of the lightest neutral electroweak gauge KK mode,
referred to here as a $\zp$, in ``Little Randall-Sundrum" (``LRS")
models \cite{LRS}, via easy-to-establish and clean final states .  
These models are volume truncations of the original RS background
and are characterized by UV-brane scales $M_5 \ll M_P$.  One can
still generate a natural hierarchy between the weak scale and the
sub-Planckian UV scale, leading to warped KK modes at the TeV scale.
However, as shown in Ref.~\cite{LRS}, several contributions to
precision data can be suppressed because of the truncation.  For
example, by taking $M_5 \sim 10^3$~TeV, corresponding to $k r_c \pi
\approx 6$, one can construct a model of flavor that avoids many of
the usual constraints and enjoys a stable hierarchy, as a result of
5D warping \cite{LRS}. A remarkable aspect of these models is the
enhancement of certain clean signals due to truncation. Simply put,
the smaller 5D volume enhances the couplings of gauge KK modes to
light fermions, while reducing the coupling strength to heavy
IR-localized states, like top quarks.  One can then show that the
typical cross section for $Z^\prime \to \ell^+ \ell^-$, with
$\ell=e,\mu$ can be enhanced as the {\it third} power of the
truncation ratio 
\beq 
y\equiv   \frac{[k r_c]_{RS}}{[k r_c]_{LRS}}\,. 
\label{y} 
\eeq

A more careful analysis of flavor constraints typically requires
somewhat larger UV scales $M_5\gsim 10^3$~TeV \cite{LRSepsK} than
originally adopted in Ref.~\cite{LRS}~\cite{BBS_LR}.
Nonetheless, it is clear that the above signal enhancement as a
function of $y>1$ can be quite  significant, {\it even for
modest truncations}, corresponding to $M_5 \gg 10^3$~TeV.
Here, we emphasize that the detection of warped KK modes by itself
is not necessarily evidence for resolving the ``Planck-weak" hierarchy, with a 5D UV scale
$M_5\sim M_P$.
In fact we believe that the experimental data needs to be used to establish
the 5D UV scale of the RS background.
The aforementioned acute sensitivity of the
$Z^\prime \to \ell^+ \ell^-$
signal to the truncation ratio $y$ affords us a valuable opportunity to
achieve this goal.

Before going further, we would like to mention that the prospects
for ``Little" KK gluons are also enhanced by the LRS
truncation~\cite{little}.  
The best final state for discovering the KK gluon will continue to be $t {\bar t}$
given that other sub-leading final states suffer from large
irreducible backgrounds.  However, 
the complications involved with $t {\bar t}$ event reconstruction makes
this channel an unlikely early
path towards establishment of a KK discovery. We will
also examine the utility of forgoing a fully reconstructed resonance
and looking for early hints of the Little KK gluon in a
lepton-counting measurement.

\underline{\it Setup}:   We will adopt the bulk gauge group
$G_B\equiv SU(2)_L \times SU(2)_R \times U(1)_X$, so that our
4D effective theory has a custodial symmetry protection and could accommodate KK modes
of mass in the 2-3~TeV range.  We note that there may be more severe bounds from flavor data
that could push the KK masses to significantly higher values \cite{Csaki:2008zd}; these considerations
may require added assumptions \cite{Fitzpatrick:2007sa,Fitzpatrick:2008zza}
and structures \cite{Csaki:2008eh}, in order to arrive at a realistic model.
Here, the gauge group $G_B$ and the  basic form of couplings among the
various states are the same as in Ref.~\cite{Zprime}.
Given that truncation does not change the structure of the 4D effective Lagrangian, we refer the
interested reader to Ref.~\cite{Zprime} for further details.  We only note that, in warped models,
the couplings of KK modes to various UV- and IR-localized states depend on the bulk volume
parameter $k r_c \pi$, and thus we could use the known results scaled by the appropriate powers
of $y$.  Specifically, the coupling of gauge KK modes to UV- (IR-) localized fermions, in a warped model,
is suppressed (enhanced) by $1/\sqrt{k r_c \pi}$ ($\sqrt{k r_c \pi}$),
compared to the SM coupling.  We will not specify a particular flavor model and will
use the approximate formulas for the coupling of gauge KK
modes to fermions~\cite{Zprime}, presented in the
appendix.  We note that a more careful analysis, based on realistic models of flavor,
could modestly change some of our results.

\underline{$Z' \rightarrow \ell^+ \ell^-$}:  
We will start by giving
simple estimates of the improvement in the LHC discovery
reach that can be attained upon truncation to LRS models \cite{LRS}.  Based on the above
discussion, the partial width $\Gamma^{\zp}_f$ for
$\zp \to f {\bar f}$, where $f=e, q,\ldots$ is a UV-localized (light)
fermion, is enhanced as $y$.  However, the total
width $\Gamma^{\zp}_T$ of the $\zp$ is still mostly controlled by the coupling to the heavy IR-localized states,
such as $W^\pm_L$ and $t$.  In this case, we expect $\Gamma^{\zp}_T $ to be
reduced by $1/y$ after truncation.
This means that the branching ratio
\beq
{\rm Br}(\zp\to \ell^+ \ell^-) \to y^2 \,{\rm Br}(\zp\to \ell^+ \ell^-),
\label{Br}
\eeq
with $\ell=e,\mu$, under truncation.  Using the narrow width approximation,
the Drell-Yan process ${\bar q}q \to \zp \to  \ell^+ \ell^-$ has a cross section
\beq
\sigma_{\rm DY} \propto \Gamma^{\zp}_q \,{\rm Br}(\zp\to \ell^+ \ell^-).
\label{prodcs}
\eeq
Using the above scaling arguments, we then see that
\beq
\sigma_{\rm DY}\to y^3 \sigma_{\rm DY},
\label{sigscaling}
\eeq
under truncation and could be significantly larger in LRS models.
We also note that the relevant SM background under the $\zp$ peak shrinks
as $1/y$, given the scaling of $\Gamma^{\zp}_T$.  Hence, we expect that
\beq
S \to y^3 S \quad; \quad S/B \to y^4 S/B,
\label{SB}
\eeq
after truncation to an LRS model \cite{LRS}.  We will next show that
the above estimates are generally confirmed by more detailed numerical
calculations of the experimental reach for the Little $\zp$.
%%%%%%%%%%%%%%%%%%%%%%%%%%%%%%%%%%%
\begin{figure}
[t]
\includegraphics[width=0.48\textwidth]{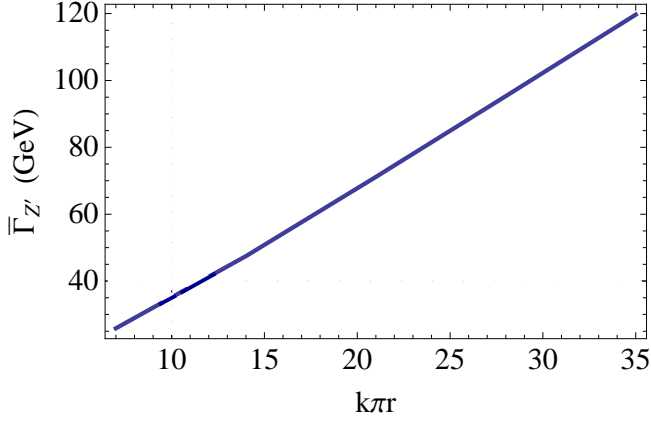}
\caption{The total width of a 2~TeV $Z'$ averaged over the three neutral states.}
\label{M2AvgGamk.FIG}
\end{figure}
%%%%%%%%%%%%%%%%%%%%%%%%%%%%%%%%
\begin{figure}
[t]
\includegraphics[width=0.48\textwidth]{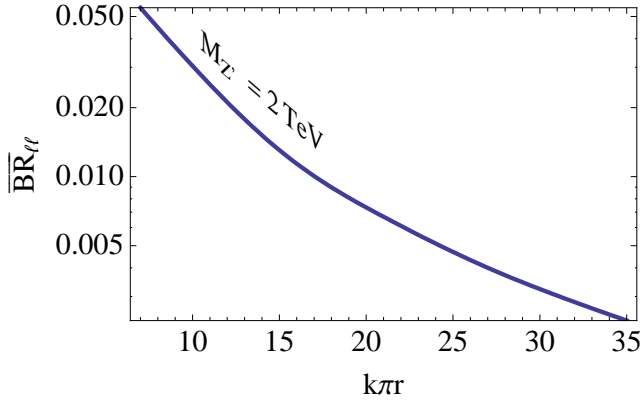}
\caption{The leptonic BR (into each of $e$ or $\mu$ pairs) of a 2~TeV $Z'$ averaged
over the three neutral states.}
\label{M2AvgBRllk.fig}
\end{figure}
%%%%%%%%%%%%%%%%%%%%%%%%%%%
\begin{figure}
[t]
\includegraphics[width=0.48\textwidth]{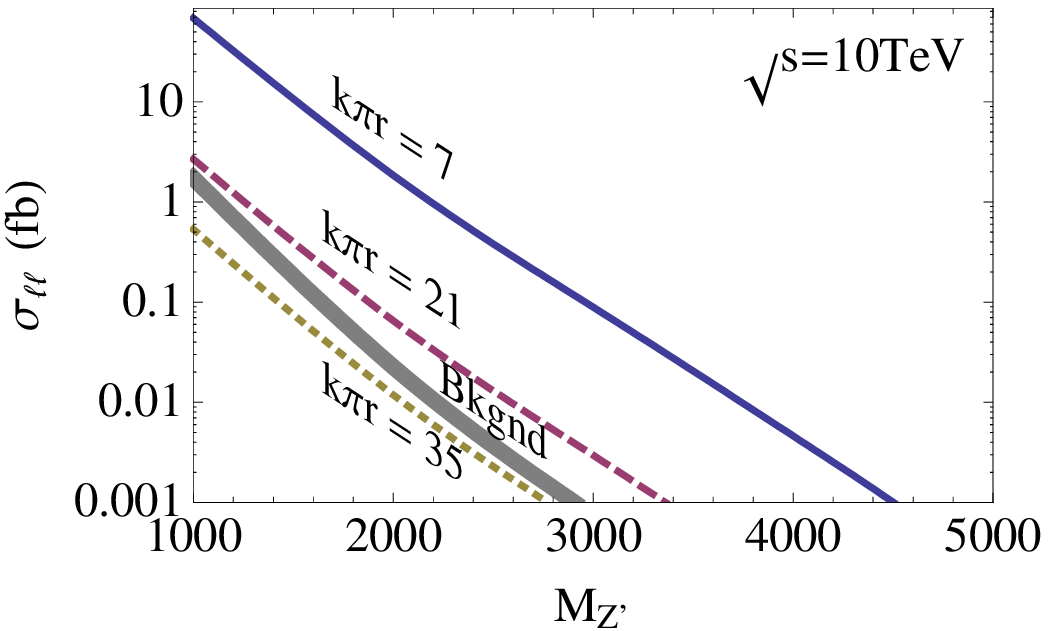}
%\caption{}
%\label{S10kxcScBM.fig}
%\end{figure}
%\begin{figure}
%%%%%%%%%%%%%%%%%%%%%%%%%%%%
\includegraphics[width=0.48\textwidth]{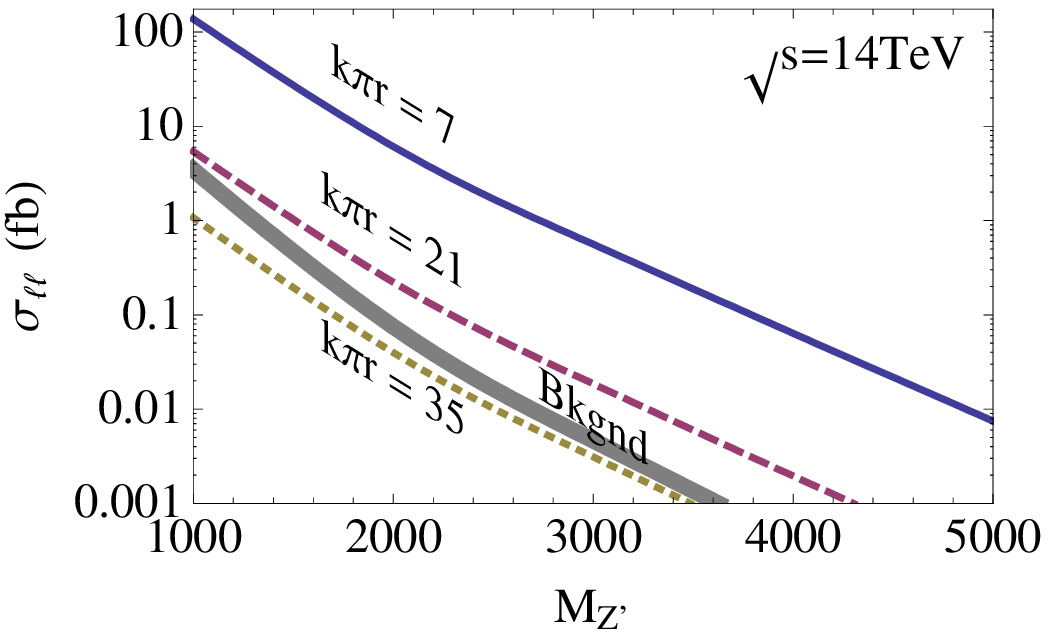}
\caption{The cross section $\sigma(pp\rightarrow Z' \rightarrow \ell^+ \ell^-)$ versus
$M_{Z'}$ at the LHC,
after the cuts in Eq.~(\ref{cuts.EQ})
(for $\ell=e$ or $\mu$, not the sum), and the SM background.  The
upper (lower) panel corresponds to $\sqrt{s}=10\, (14)$~TeV.}
\label{SxkxcScBM.fig}
\end{figure}
\begin{figure}
[t]
%%%%%%%%%%%%%%%%%%
\includegraphics[width=0.48\textwidth]{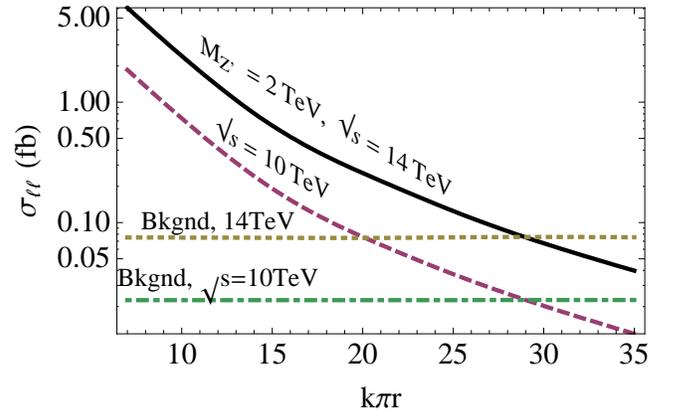}
\caption{The predicted $\sigma(pp\rightarrow Z' \rightarrow \ell^+ \ell^-)$ at the LHC
after the cuts in Eq.~(\ref{cuts.EQ})
(for $\ell=e$ or $\mu$, not the sum), as a function of $\krp$, and the SM background.  Here,
we have set $M_{Z'}=2$~TeV.}
\label{SxM2cScBk.FIG}
\end{figure}
\begin{figure}
[t]
%%%%%%%%%%%%%%%%%%%%%%%%%%%%%%
\includegraphics[width=0.48\textwidth]{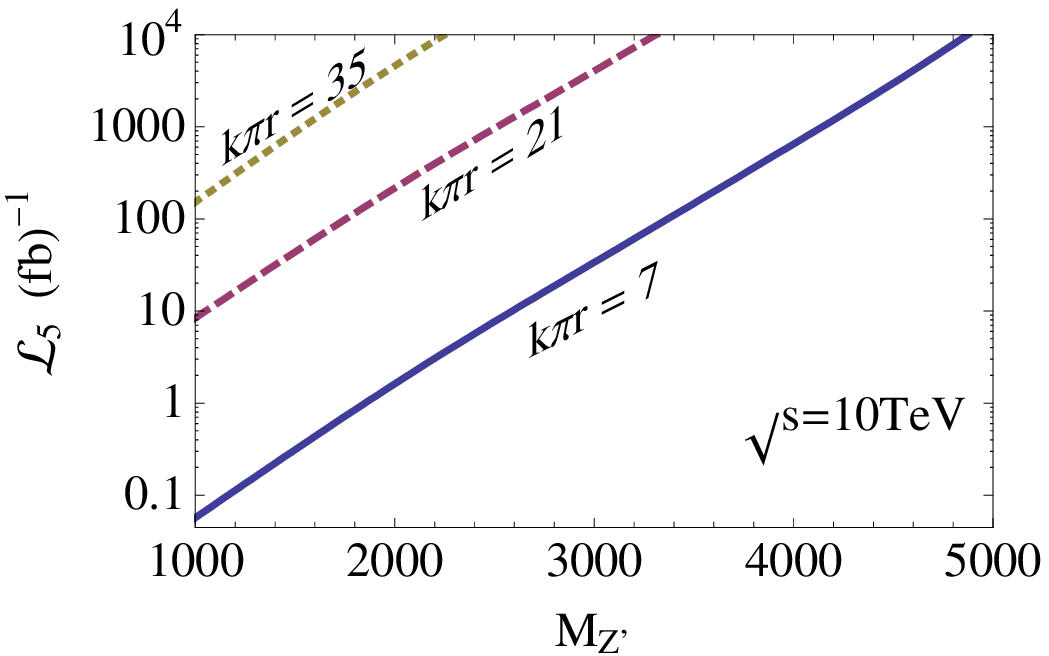}
%\caption{}
%\label{fig4}
%\end{figure}
%\begin{figure}
%[t]
\includegraphics[width=0.48\textwidth]{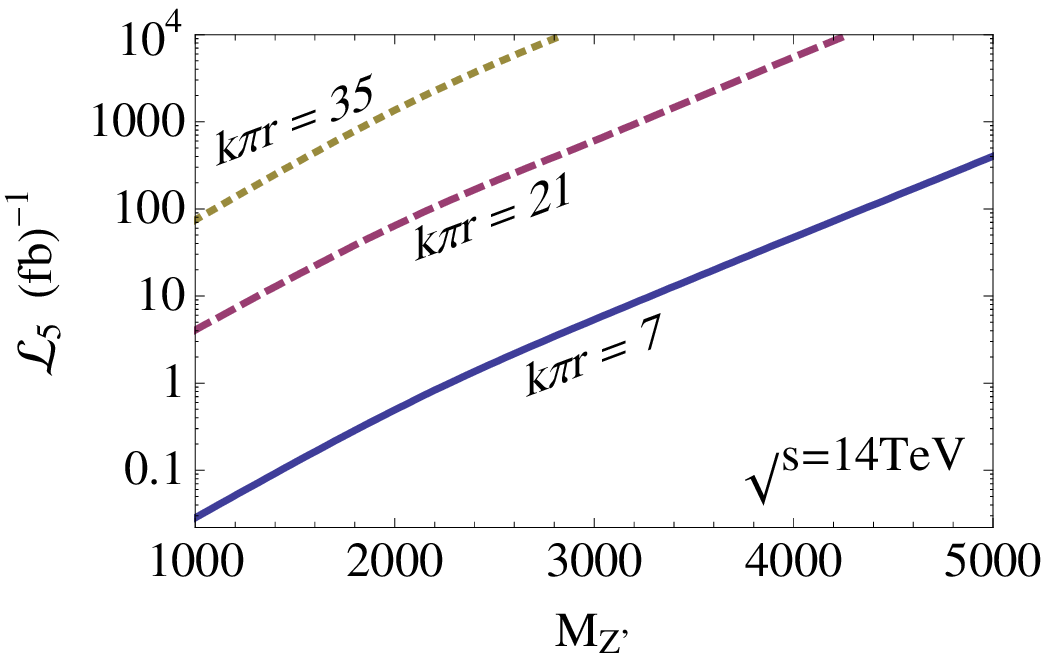}
\caption{The luminosity required for $5\, \sigma$ significance at the LHC
in the $pp \rightarrow \ell^+\ell^-$ channel (with $\ell=e$ or $\mu$, not the sum)
requiring at least 3 events.  The
upper (lower) panel corresponds to $\sqrt{s}=10\, (14)$~TeV.}
\label{SxkxL5PGM3Ev.FIG}
\end{figure}
%%%%%%%%%%%%%%%%%%%%%%%%%%%%%%%%%%%%%%%
Figs.~\ref{M2AvgGamk.FIG} and \ref{M2AvgBRllk.fig},
respectively show the total width and the leptonic Br of a 2-TeV $Z'$,
averaged over the three neutral states, as a function of $\krp$;
the Br shown is into each lepton ($e$ or $\mu$, not the sum).
These plots confirm the scaling suggested by the preceding analysis based on the
behavior of the couplings as $\krp$ is varied.
As expected, the width and the leptonic Br have approximate
$1/y$ and $y^2$ scaling, respectively, under truncation.

At the LHC, to maximize the $\zp$ signal above the SM background,
we apply the cuts
\begin{eqnarray}
|\eta_\ell| < 3.0 \ \ ,& & \ \ {p_T}_\ell > 100~{\rm GeV} \ \ \rm{and} \nonumber \\
(M_{Z'}-100~{\rm GeV}) < & & \!\!\!\!\!\! M_{\ell^+\ell^-} \!\!< (M_{Z'}+100~{\rm GeV}).
\label{cuts.EQ}
\end{eqnarray}
In Fig.~\ref{SxkxcScBM.fig}, we present
$\sigma(pp\rightarrow Z' \rightarrow \ell^+ \ell^-)$ and the SM background
(for $\ell=e$ or $\mu$, not the sum) versus $M_{Z'}$, at $\sqrt{s}=10\,(14)$~TeV,
for $k\pi r = 7,21,35$, in the upper (lower) panel.  Fig.~\ref{SxM2cScBk.FIG} again
shows the leptonic cross section for
$M_{Z'}=2$~TeV, after cuts, but as a function of $\krp$.
Given the sufficiently low level ($\sim 10^{-3}$)
of leptonic jet-fakes expected \cite{jetfakes},
we have included only the
SM irreducible backgrounds.
As can be deduced from these figures, for $M_{Z'} \gsim 2$~TeV, once
$\krp \lsim 20$, corresponding to $y \gsim 2$, we find that $S/B\gg1$ and
the measurements become essentially background free.

The luminosity required
for a $5\,\sigma$ discovery of the
$Z'$ at the LHC, as a function of $M_{Z'}$, is
given in Fig.~\ref{SxkxL5PGM3Ev.FIG}.  We consider
$\sqrt{s}=10 (14)$~TeV, in the upper (lower) panel, and $\krp =7,21,35$.
In determining the luminosity required, we require at least 3 events be observed,
and if the number of events is small we use Poisson statistics
to compute the significance equivalent to $5\sigma$.
Given that $S/B\gg 1$
for parameters of interest to our analysis, these events are basically
background-free, as mentioned before.
Hence, the use of only a handful of events
to establish a discovery is reasonable.

We can see from the upper panel in Fig.~\ref{SxkxL5PGM3Ev.FIG}
that even at $\sqrt{s}=10$~TeV,
as may be the case during the initial phase of the LHC operation, one can begin
to place meaningful bounds on $\krp$, for $M_{Z'}\approx 2$~TeV, with a modest
integrated luminosity of about $1$~fb$^{-1}$.  At $\sqrt{s}=14$~TeV,
a 5$\sigma$ discovery of a 3-TeV $\zp$ requires about \fb{4}.
Note that the plots represent
only one leptonic decay channel and the final reach is attained by
summing over both $e^\pm$ and $\mu^\pm$ channels.
The lower panel of the same figure shows that,
assuming $\krp\approx 7$ and $\sqrt{s}=14$~TeV,
the reach for the $M_{Z'}$ will be about 5~TeV, with
$\ord{300}$~fb$^{-1}$.

Our results demonstrate that,
for phenomenologically viable parameters,
the reach for the Little $Z'$ in clean di-lepton channels is
greatly enhanced in truncated models.  By contrast, as can also be
seen from these plots, for $M_{Z'}\gsim 2$ TeV, in models
that explain the Planck-weak hierarchy ($\krp \approx 35$) an integrated luminosity
of order 1000~fb$^{-1}$ is required.  Even utilizing other
decay channels, such as $W_L^\pm$, discovery at $M_{Z'}\sim 3$~TeV
in the RS model will require
upwards of $\ord{300}$~fb$^{-1}$ \cite{Zprime}.

\underline{\it Level-2 KK modes}: Given the much improved discovery reach for the $Z'$
in truncated LRS models, we would like to examine the possibility of
detecting the second level of resonances in the KK tower,
collectively denoted by $Z''$, in the clean di-lepton channel.
Discovering the $Z''$ allows one to confirm the
prediction for the ratio $M_{Z'}/M_{Z''}$ in warped models (which is
different from, say,  those in models with flat extra dimensions).  We note that
the prospects for achieving this in the full RS model with $\krp \approx 35$
is practically nil.

To estimate the reach for $Z''$ in LRS models, we
take $\krp = 7$, near the lower bound allowed for a warped model of
flavor largely unconstrained by precision data.  For this point in the
parameter space, we have $M_{Z'}/M_{Z''} \simeq 2.65/5.80\simeq 0.46$;
the prediction
in the RS model with $\krp =35$ is
$M_{Z'}/M_{Z''} \simeq 2.45/5.57\simeq 0.44$.  Such a
difference in mass ratios is expected to
be measurable, as the clean $e^\pm$ final state is quite
accessible for this LRS example.
We have checked that the coupling of the $Z''$
to the light SM fermions will be smaller by the ratio
$\sim 0.14/0.18$, compared
to those of the $Z'$.  This can be easily
verified using typical wavefunctions for fermion zero modes
in our LRS model \cite{LRS}.

Using the set of cuts (\ref{cuts.EQ}) and setting
$M_{Z'}=2$~TeV, we find that the Little $Z''$
with $M_{Z''}\simeq 4.3$~TeV can be
discovered, with $\ord{100}$~fb$^{-1}$
at $\sqrt{s}=14$~TeV.
Hence, for $\krp\approx 7$, compatible with
the requirements of a viable flavor scenario, we see that the
second KK resonance can be discovered.
This will in turn provide further information
on the warped nature of the underlying geometry.

We close this discussion by noting that we have limited our analysis
to the effects of UV-truncations in this work.  In principle, one could also
imagine deforming the RS background in the IR, by adding
brane-localized gauge kinetic terms, resulting in modulations
of gauge KK masses and their couplings
to the IR-localized fields \cite{Davoudiasl:2002ua,Carena:2002dz}.
We have implicitly assumed that brane kinetic terms
are generated at quantum level and are small.  However, one could also study the
effects of such terms on collider phenomenology \cite{Lillie:2007ve}.  These brane terms
generally change the relation between KK masses and couplings in a way that
is quite different from that resulting from truncation.  For example, the
IR-brane kinetic terms required to get
$\ord{1}$ reductions in the gauge KK couplings to the IR-brane yield
$M^{(1)}/M^{(2)} \lsim 1/3$ for the ratio of the first two
KK masses \cite{Davoudiasl:2002ua,Carena:2002dz}.  As we have seen,
this is not the case in the truncated models examined in this work.

\underline{\it $g^{(1)} \rightarrow t \bar t$}:
Finally, we comment on the expected enhancements in the reach
for the KK gluon
in truncated flavor models.  In particular,
the lightest gluon KK mode $g^{(1)}$ in warped models generally has the largest
production rate \cite{KKgluon1,KKgluon2}.  Here, the same enhanced
coupling to light fermions as for the $Z'$ would also lead
to an enhanced production rate
for $g^{(1)}$.  
However, although the decay into light fermions would also be enhanced,
the di-quark final state is swamped by a large di-jet background,
and since the KK gluon does not directly couple to leptons,
no clean di-lepton final state exists.
%$g^{(1)}$ does not have a
%``clean" light fermion final state
% and also due to the large di-jet background from QCD.
Thus, one would use the $t {\bar t}$ final state to look for Little KK gluons of the LRS model,
as in the RS model with $\krp \approx 35$
\cite{KKgluon1,KKgluon2}.  Using the couplings presented in the appendix,
we find the width of  $g^{(1)}$ into $t {\bar t}$, Br$(g^{(1)}\to t {\bar t}) \sim 1$,
over most of the relevant parameter space.   Hence, the predicted width
of Little KK gluons
will decrease roughly as $1/y$.
Parametrically, we then predict $S \sim y$, $S/\sqrt{B} \sim y/^{3/2}$, and
$S/B \sim y^2$ for $pp \to g^{(1)} \to t {\bar t}$.
This behavior can significantly extend the reach for
Little KK gluons, well above KK masses ($\sim 4$~TeV \cite{KKgluon1,KKgluon2})
accessible at the LHC with $\ord{10^2}$~fb$^{-1}$
in untruncated models.  In addition,
one may expect that the enhanced
Little $g^{(1)}$ production could provide very
early hints of new physics at the LHC.

Before a detailed numerical analysis,
to get an estimate of the early Little $g^{(1)}$ signal, we will
use the results of Ref.~\cite{KKgluon1} that imply a 3-TeV KK gluon can be
detected at the $5\sigma$ level with about \fb{25}
of integrated luminosity\footnote{However, we note that further refinement of
this analysis may be possible, using a more detailed understanding
of top-jet morphology \cite{topjet}.}.
On the other hand,
our Fig.~\ref{SxkxcScBM.fig} roughly suggests that for $\krp \approx 7$
we could expect a production cross-section enhancement of about 20 in going
from 3~TeV to 2~TeV in the KK mass, whereas going from $\sqrt{s}=14$~TeV to
$\sqrt{s}=10$~TeV reduces the cross section by a factor of about 4 
($BR_{\ell\ell}$ doesn't change much in these translations, and
can be factored out of Fig.~\ref{SxkxcScBM.fig}).
We also note that  for $k r_c\pi \approx 7$ we get Br$(g^{(1)}\to t {\bar t}) \approx 1/2$, using the
rough expressions in the appendix.  Given these
considerations, for $y\approx 5$ ($k r_c\pi \approx 7$), we estimate that a
$5\sigma$ signal for a 2-TeV Little KK gluon only requires
about 300~pb$^{-1}$ at $\sqrt{s}=10$~TeV (early LHC operation), whereas a 3-TeV
state would require roughly 2~fb$^{-1}$ at $\sqrt{s}=14$~TeV.

The preceding analysis, on the first
look, seems to suggest that we may be able to establish the RS-type
properties of the Little KK gluon
somewhat ahead of the Little $\zp$.
However, we note that the results of Ref.~\cite{KKgluon1} are based
on cuts involving missing transverse momenta from the leptonic decays of one of the
$W$ bosons in the top decay chain, and reconstructing the top.
Such measurements may not be well-understood
at the early stages of the LHC experiments.
One could instead focus on the leptonic
decays of both $W$ bosons, coming from the decays  of the $\bar t t$
pair, to see if the di-lepton signal is large enough compared
to the background to afford an early hint of new physics.
The price to pay
is the branching fraction of $W$ into $e$ or $\mu$ plus $\nu$, which is $2/9$,
and our inablility to form the highly effective $t\bar t$ invariant mass since
the event cannot be fully reconstructed due to the presence of two missing neutrinos.
We will look at this possibility more carefully below.  We will begin with a more
accurate numerical assessment of the
preceding scaling estimates for KK gluon detection.

%%%%%%%%%%%%%%%%%%%%%%%%%%%%%%%%%%%%%%%%%%
%%%%%%%%%%%%%%%%%%%%%%%%%%%%%%%%%%%%%%%%%%

To find roughly what the LHC reach for the KK gluon is, we have implemented its couplings
in the Monte Carlo package CalcHEP~\cite{CalcHEP}.
For a 3~TeV $g^{(1)}$ we find that the total width in GeV
(Br$_{t \bar t}$ in parentheses) is
71 (0.55), 196 (0.9), 344 (0.93) for $k r_c\pi =7,~21,~35$ respectively.
Since the Br into $t\bar t$ is large and the backgrounds can be brought under
control, we will focus on this decay mode.

\begin{figure}
\includegraphics[width=0.35\textwidth,angle=270]{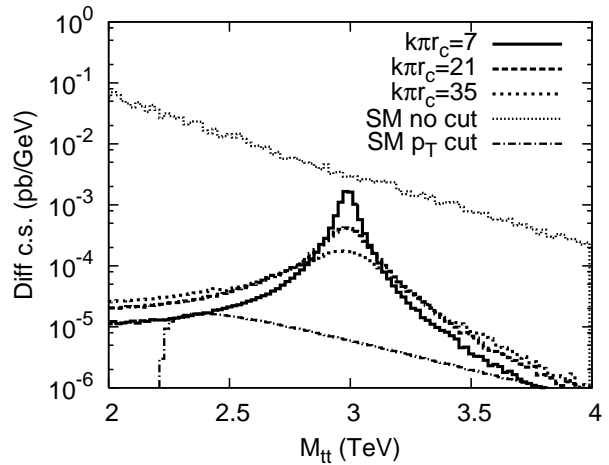}
\caption{
The $pp\rightarrow t \bar t$ invariant mass distribution from production
and decays  of Little $g^{(1)}$ and SM background including a
$p_T$ cut of 1100 GeV is shown. 
}
\label{Mtt3TeV.FIG}
\end{figure}
In Fig.~\ref{Mtt3TeV.FIG}, we show the $t\bar t$ invariant mass distribution
from $pp\rightarrow g^{(1)}\rightarrow t \bar t$ without any top decay BRs included.
We show the signal and irreducible SM background without any cuts,
and in addition, also the SM background curve with ${p_T}_t > 1100$~GeV cut.
If one can reconstruct the tops fully, we see that with the large $p_T$ cut
${p_T}_t > 1100$~GeV, the signal is comfortably above background.
Note that in this plot, the signal is not shown after the cut,
and the total cross section is reduced only by about a factor of two (three)
for $k\pi r = 7$ ($35$) after the cut, as we can infer from Table~\ref{pp2KKG2tT.TAB}
which shows the total cross section before and after cuts.
\begin{table}
\begin{center}
\caption{The total cross section for $p p \to t \bar t$ in fb before and after the cuts
${p_T}_t > 1100$~GeV, $|y_t| < 3$.
The $g^{(1)}$ signal and irreducible SM background are shown.
\label{pp2KKG2tT.TAB}}
\begin{tabular}{|c|c|c|c|c|}
\hline
$\sigma_{t\bar{t}}$ (fb) &
$k r_c\pi =7$&
$k r_c \pi =21$&
$k r_c\pi =35$&
SM\tabularnewline
\hline
\hline
No cuts&
$197$&
$127$&
$91$&
$4\times10^{5}$\tabularnewline
\hline
$p_{T_{t}}$, $y_{t}$ cuts&
$100$&
$52$&
$30$&
$11$\tabularnewline
\hline
\end{tabular}
\end{center}
\end{table}
Since reconstructing the (boosted) top is not completely achievable in a hadron
collider, this clear signal peak will be somewhat degraded after taking into account
top decays.
We will discuss next the LHC signals for specific decay channels of the top.

In the semi-leptonic channel one can reconstruct the $t \bar t$ event in the
transverse plane.
Applying $W$ and top mass constraints, one can even fully reconstruct the event.
However, the large boost of the tops will bring in additional
complications. We will leave a detailed analysis of this channel to future work.
Here, we will perform a simple-minded estimate for the reach in the semi-leptonic channel.
Assuming a 5\% efficiency~\cite{KKgluon1} for reconstructing a hadronic top
(which includes $b$-tagging efficiency and kinematic acceptances),
we find, for a 3~TeV KK gluon, that the integrated luminosity needed for a
$5\sigma$ discovery at the LHC is
about $2,~8,~21~{\rm fb^{-1}}$ for $k r_c \pi =7,~21,~35$ respectively.
Here, we will content ourselves with this rough estimate for the semi-leptonic mode,
but we note that we have good agreement for the $k r_c\pi=35$ case with the conclusions
of the more complete study in Ref.~\cite{KKgluon1}, and for the $k r_c \pi=7$ case with
our scaling estimate.

\begin{figure}
\includegraphics[width=0.35\textwidth,angle=270]{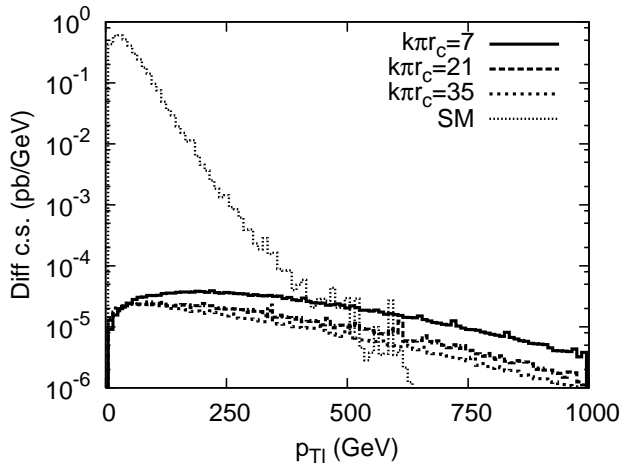}
\caption{
The lepton $p_T$ distribution in the leptonic decay channel of
the top in $p p \rightarrow t \bar t$, after the basic cuts $|\eta_{\ell, b}| < 3$.
The signal is for a 3~TeV KK gluon.
\label{pTl3TeV.FIG} }
\end{figure}

The di-lepton channel is experimentally cleaner given that we have two hard leptons.
However, due to the two missing neutrinos, the event cannot be fully reconstructed.  
Here, we will restrict ourselves to a simple di-lepton counting
analysis.  Given that the lepton $p_T$ is correlated with the $p_T$ of its parent top,
and our observations in Fig.~\ref{Mtt3TeV.FIG}, the simplest thing one could do is
to cut hard on the lepton $p_T$ and count the di-lepton events.
To see how well this can work, we show the $p_T$ distribution of the lepton
from the top for the signal and the irreducible SM background
in Fig.~\ref{pTl3TeV.FIG}.
Based on this distribution, we apply the hard cut ${p_T}_\ell > 400$~GeV,
in addition to the other soft cuts $|\eta_{\ell,b}| < 3$, ${p_T}_b > 20$~GeV.
We find that, for a $5\sigma$ discovery at the LHC, we need an integrated luminosity
of $4,~15,~109~{\rm fb^{-1}}$ for $k r_c \pi =7,~21,~35$ respectively,
for a 3~TeV KK gluon.  
Taking into account the 2 $b$-jets that are present in the events can 
undoubtedly improve the reach,
however, our focus in this paper is on the dilepton signal.  
A full analysis must also take into account the complication of
the isolation being spoiled due to the presence of the $b$-jet close to the
lepton, owing to the large top boost.

\underline{\it Holographic interpretation}: Here, we note that,
based on the AdS/CFT correspondence \cite{Maldacena:1997re}, RS-type
models have a 4D dual interpretation, in terms of some unknown
strong dynamics \cite{ArkaniHamed:2000ds,Rattazzi:2000hs}. In the
dual 4D picture, the hierarchy is related to the dynamical emergence
of a conformal theory below the UV scale (originally taken to be
$M_5 \sim M_P \sim 10^{19}$~GeV).  Upon breaking of the conformal
symmetry an IR scale of order the weak scale appears, as signaled by
the appearance of various composite resonances (such as the KK
modes).  In LRS-type models, the conformal energy interval is
truncated in the UV and spans a smaller range. Our preceding
discussion regarding the signals of truncated models was
geometrical. However, using the duality, one can interpret our
results as providing clues on the size of the conformal interval
(corresponding to $\krp$), as well as the nature of the
corresponding 4D dynamics (duality with warping).

%%%%%%%%%%%%%%%%%%%%%%%%%%%%%%%%%%%%%%%%%
%%%%%%%%%%%%%%%%%%%%%%%%%%%%%%%%%%%%%%%%%

\underline{\it Summary}:  In this work, we considered the possibility
that the UV scale of warped 5D models could be well-below the
4D scale of gravity $M_P$.  In this case, one could still achieve a natural
though smaller hierarchy between a high UV scale and the weak scale.  These
truncated LRS backgrounds can accommodate  realistic
warped flavor models that are free from some of the constraints that exist
when the UV scale is near $M_P$.  At the same time, the truncated models
offer greatly improved LHC discovery prospects.  We concentrated on the neutral
$Z'$ resonances that arise in warped models with bulk custodial symmetry.
We found that for sufficiently truncated yet viable models ($M_5 \sim 10^4$~TeV), the LHC
discovery of a 2(3)-TeV Little $Z'$ in the experimentally clean
$e^\pm$ and $\mu^\pm$ channels may
require only about 1 (4)~fb$^{-1}$, at $\sqrt{s}=10\, (14)$~TeV.
At $\sqrt{s}=14$~TeV, with $\ord{100}$~fb$^{-1}$, the corresponding
reach is about 5~TeV, well into the region allowed by EW precision data.
For similar truncations, one may be able to discover
the second level resonance $Z''$ in the same channels.  The precision
afforded by the di-lepton final states can then result in determining the
ratio of $Z'$ and $Z''$ masses.  The sensitive dependence of the cross
section for $pp \to Z' \to \ell^+ \ell^-$ on truncation allows one to probe the
size of the hierarchy generated by the 5D gravitational redshift (4D conformal
dynamics).  If the masses of successive  resonances can be measured, as
may be the case in LRS models, the warped nature of the 5D picture can also
be corroborated.  We also considered the expected enhanced
discovery prospects for other KK modes, such as KK gluons, in
truncated models.
Significant enhancements
in the reach for warped Little $W'$ resonances \cite{Agashe:2008jb},
based on considerations similar to those presented above may also be expected.
Even though the experimental prospects for Little
KK gluons improve considerably, our analysis suggests that they do not
offer a substantial advantage for early detection, compared to the leptonically
decaying Little $\zp$ modes; this is due to our conservative assumptions
regarding the early reconstruction efficiency for the $t {\bar t}$ final state, which
will presumably improve once the experiments mature.  In any event, a
more detailed study using improved boosted top-jet reconstruction
techniques may enhance the Little KK gluon early detection prospects.

%\acknowledgments
{\it Acknowledgments:}
We thank Gilad Perez for many discussions especially pertaining to KK gluons in
LRS models.  This work was supported by the US Department of
Energy under Grant Contract DE-AC02-98CH10886.

\section{Appendix}
\appendix*

Here, we present the set of approximate expressions for the couplings of various
fermions to the gauge first KK modes, used in our analysis.  The following are
in units of of the appropriate SM gauge couplings.
With $\xi\equiv \sqrt{k r_c \pi}$, for $t_R$ and $(t, b)_L$ we take~\cite{Zprime}
\beq
-1.13/\xi + 0.7 \,\xi\; \;\;{\rm and}\;
-1.13/\xi + 0.2 \,\xi\,,
\label{tbL}
\eeq
respectively.
For all other fermions we assume
\beq
-1.13/\xi\,.
\label{lightf}
\eeq

We note that these formulas
were used only as a rough guide to the
behavior of the couplings under truncation.  However,
in principle, a more detailed study would choose different fermion profiles at different
values of $k r_c \pi$.  Sample profiles for the case $k r_c \pi \sim 7$ have been provided in
Refs.~\cite{LRS,LRSepsK}.

%%%%%%%%%%%%%%%%%%%%%%%%%%%%%%%%%%%%%%%%%%%%%%%%%%%%%%%%%%%%%%%%%%%%%%%%%%%


\begin{thebibliography}{99}

%\cite{Randall:1999ee}
\bibitem{Randall:1999ee}
  L.~Randall and R.~Sundrum,
  %``A large mass hierarchy from a small extra dimension,''
  Phys.\ Rev.\ Lett.\  {\bf 83}, 3370 (1999)
  [arXiv:hep-ph/9905221].
  %%CITATION = PRLTA,83,3370;%%

%\cite{Davoudiasl:1999jd}
\bibitem{Davoudiasl:1999jd}
  H.~Davoudiasl, J.~L.~Hewett and T.~G.~Rizzo,
  %``Phenomenology of the Randall-Sundrum gauge hierarchy model,''
  Phys.\ Rev.\ Lett.\  {\bf 84}, 2080 (2000)
  [arXiv:hep-ph/9909255].
  %%CITATION = PRLTA,84,2080;%%

%\cite{Goldberger:1999wh}
\bibitem{Goldberger:1999wh}
  W.~D.~Goldberger and M.~B.~Wise,
  %``Bulk fields in the Randall-Sundrum compactification scenario,''
  Phys.\ Rev.\  D {\bf 60}, 107505 (1999)
  [arXiv:hep-ph/9907218].
  %%CITATION = PHRVA,D60,107505;%%

%\cite{Davoudiasl:1999tf}
\bibitem{Davoudiasl:1999tf}
  H.~Davoudiasl, J.~L.~Hewett and T.~G.~Rizzo,
  %``Bulk gauge fields in the Randall-Sundrum model,''
  Phys.\ Lett.\  B {\bf 473}, 43 (2000)
  [arXiv:hep-ph/9911262].
  %%CITATION = PHLTA,B473,43;%%

%\cite{Pomarol:1999ad}
\bibitem{Pomarol:1999ad}
  A.~Pomarol,
  %``Gauge bosons in a five-dimensional theory with localized gravity,''
  Phys.\ Lett.\  B {\bf 486}, 153 (2000)
  [arXiv:hep-ph/9911294].
  %%CITATION = PHLTA,B486,153;%%

%\cite{Grossman:1999ra}
\bibitem{Grossman:1999ra}
  Y.~Grossman and M.~Neubert,
  %``Neutrino masses and mixings in non-factorizable geometry,''
  Phys.\ Lett.\  B {\bf 474}, 361 (2000)
  [arXiv:hep-ph/9912408].
  %%CITATION = PHLTA,B474,361;%%

%\cite{Gherghetta:2000qt}
\bibitem{Gherghetta:2000qt}
  T.~Gherghetta and A.~Pomarol,
  %``Bulk fields and supersymmetry in a slice of AdS,''
  Nucl.\ Phys.\  B {\bf 586} (2000) 141
  [arXiv:hep-ph/0003129].
  %%CITATION = NUPHA,B586,141;%%

%\cite{Agashe:2003zs}
\bibitem{custodial}
  K.~Agashe, A.~Delgado, M.~J.~May and R.~Sundrum,
  %``RS1, custodial isospin and precision tests,''
  JHEP {\bf 0308}, 050 (2003)
  [arXiv:hep-ph/0308036];
  %%CITATION = JHEPA,0308,050;%%
%\cite{Agashe:2006at}
%\bibitem{Agashe:2006at}
  K.~Agashe, R.~Contino, L.~Da Rold and A.~Pomarol,
  %``A custodial symmetry for Z b anti-b,''
  Phys.\ Lett.\  B {\bf 641}, 62 (2006)
  [arXiv:hep-ph/0605341].
  %%CITATION = PHLTA,B641,62;%%

%\cite{Carena:2007ua}
\bibitem{Carena:2007ua}
  M.~S.~Carena, E.~Ponton, J.~Santiago and C.~E.~M.~Wagner,
  %``Electroweak constraints on warped models with custodial symmetry,''
  Phys.\ Rev.\  D {\bf 76}, 035006 (2007)
  [arXiv:hep-ph/0701055].
  %%CITATION = PHRVA,D76,035006;%%

%\cite{Agashe:2006hk}
\bibitem{KKgluon1}
  K.~Agashe, A.~Belyaev, T.~Krupovnickas, G.~Perez and J.~Virzi,
  %``LHC signals from warped extra dimensions,''
  arXiv:hep-ph/0612015.
  %%CITATION = HEP-PH/0612015;%%

%\cite{Lillie:2007yh}
\bibitem{KKgluon2}
  B.~Lillie, L.~Randall and L.~T.~Wang,
  %``The Bulk RS KK-gluon at the LHC,''
  arXiv:hep-ph/0701166.
  %%CITATION = HEP-PH/0701166;%%

%\cite{Agashe:2007ki}
\bibitem{Zprime}
  K.~Agashe {\it et al.},
  %``LHC Signals for Warped Electroweak Neutral Gauge Bosons,''
  Phys.\ Rev.\  D {\bf 76}, 115015 (2007)
  [arXiv:0709.0007 [hep-ph]].
  %%CITATION = PHRVA,D76,115015;%%


%\cite{Agashe:2008jb}
\bibitem{Agashe:2008jb}
  K.~Agashe, S.~Gopalakrishna, T.~Han, G.~Y.~Huang and A.~Soni,
  %``LHC Signals for Warped Electroweak Charged Gauge Bosons,''
  arXiv:0810.1497 [hep-ph].
  %%CITATION = ARXIV:0810.1497;%%

%\cite{Fitzpatrick:2007qr}
\bibitem{KKgraviton1}
  A.~L.~Fitzpatrick, J.~Kaplan, L.~Randall and L.~T.~Wang,
  %``Searching for the Kaluza-Klein graviton in bulk RS models,''
  arXiv:hep-ph/0701150.
  %%CITATION = HEP-PH/0701150;%%

%\cite{Agashe:2007zd}
\bibitem{KKgraviton2}
  K.~Agashe, H.~Davoudiasl, G.~Perez and A.~Soni,
  %``Warped Gravitons at the LHC and Beyond,''
  Phys.\ Rev.\  D {\bf 76}, 036006 (2007)
  [arXiv:hep-ph/0701186].
  %%CITATION = PHRVA,D76,036006;%%

%\cite{Djouadi:2007eg}
\bibitem{LHC-KK}
For some additional works on the collider phenomenology
of warped models see,
  A.~Djouadi, G.~Moreau and R.~K.~Singh,
  %``Kaluza--Klein excitations of gauge bosons at the LHC,''
  Nucl.\ Phys.\  B {\bf 797}, 1 (2008)
  [arXiv:0706.4191 [hep-ph]];
  %%CITATION = NUPHA,B797,1;%%
%\cite{Antipin:2007pi}
%\bibitem{Antipin:2007pi}
  O.~Antipin, D.~Atwood and A.~Soni,
  %``Search for RS gravitons via $W_L W_L$ decays,''
  Phys.\ Lett.\  B {\bf 666}, 155 (2008)
  [arXiv:0711.3175 [hep-ph]].
  %%CITATION = PHLTA,B666,155;%%

%\cite{Davoudiasl:2008hx}
\bibitem{LRS}
  H.~Davoudiasl, G.~Perez and A.~Soni,
  %``The Little Randall-Sundrum Model at the Large Hadron Collider,''
  Phys.\ Lett.\  B {\bf 665}, 67 (2008)
  [arXiv:0802.0203 [hep-ph]].
  %%CITATION = PHLTA,B665,67;%%

%\cite{Bauer:2008xb}
\bibitem{LRSepsK}
  M.~Bauer, S.~Casagrande, L.~Grunder, U.~Haisch and M.~Neubert,
  %``Little Randall-Sundrum models: epsilon_K strikes again,''
  arXiv:0811.3678 [hep-ph].
  %%CITATION = ARXIV:0811.3678;%%

\bibitem{BBS_LR}
This is a consequence
mainly of $K-\bar K$
constraints resulting in large part from enhanced matrix elements
of LR operators, see
G. Beall, M. Bander and A. Soni,
Phys.\ Rev.\ Lett.\  {\bf 48}, 848 (1982); see also M. Bona {\it et al}
[UTfit
Collab.],  JHEP {\bf 0803}, 049 (2008)

\bibitem{little}
All Kaluza-Klein excitations  
resulting from any truncation different from the original RS value of
 $k r_c \pi \approx 35$ will be termed ``Little". 

%\cite{Csaki:2008zd}
\bibitem{Csaki:2008zd}
  C.~Csaki, A.~Falkowski and A.~Weiler,
  %``The Flavor of the Composite Pseudo-Goldstone Higgs,''
  JHEP {\bf 0809}, 008 (2008)
  [arXiv:0804.1954 [hep-ph]].
  %%CITATION = JHEPA,0809,008;%%

%\cite{Fitzpatrick:2007sa}
\bibitem{Fitzpatrick:2007sa}
  A.~L.~Fitzpatrick, G.~Perez and L.~Randall,
  %``Flavor from Minimal Flavor Violation & a Viable Randall-Sundrum Model,''
  arXiv:0710.1869 [hep-ph].
  %%CITATION = ARXIV:0710.1869;%%

%\cite{Fitzpatrick:2008zza}
\bibitem{Fitzpatrick:2008zza}
  A.~L.~Fitzpatrick, L.~Randall and G.~Perez,
  %``Flavor Anarchy In A Randall-Sundrum Model With 5d Minimal Flavor Violation
  %And A Low Kaluza-Klein Scale,''
  Phys.\ Rev.\ Lett.\  {\bf 100}, 171604 (2008).
  %%CITATION = PRLTA,100,171604;%%

%\cite{Csaki:2008eh}
\bibitem{Csaki:2008eh}
  C.~Csaki, A.~Falkowski and A.~Weiler,
  %``A Simple Flavor Protection for RS,''
  arXiv:0806.3757 [hep-ph].
  %%CITATION = ARXIV:0806.3757;%%

\bibitem{jetfakes}
G.~Wooden, ``Leptonic fake rates from jets at ATLAS," ATL-PHYS-SLIDE-2009-078.

%\cite{Davoudiasl:2002ua}
\bibitem{Davoudiasl:2002ua}
  H.~Davoudiasl, J.~L.~Hewett and T.~G.~Rizzo,
  %``Brane localized kinetic terms in the Randall-Sundrum model,''
  Phys.\ Rev.\  D {\bf 68}, 045002 (2003)
  [arXiv:hep-ph/0212279].
  %%CITATION = PHRVA,D68,045002;%%

%\cite{Carena:2002dz}
\bibitem{Carena:2002dz}
  M.~S.~Carena, E.~Ponton, T.~M.~P.~Tait and C.~E.~M.~Wagner,
  %``Opaque branes in warped backgrounds,''
  Phys.\ Rev.\  D {\bf 67}, 096006 (2003)
  [arXiv:hep-ph/0212307].
  %%CITATION = PHRVA,D67,096006;%%

%\cite{Lillie:2007ve}
\bibitem{Lillie:2007ve}
  B.~Lillie, J.~Shu and T.~M.~P.~Tait,
  %``Kaluza-Klein Gluons as a Diagnostic of Warped Models,''
  Phys.\ Rev.\  D {\bf 76}, 115016 (2007)
  [arXiv:0706.3960 [hep-ph]].
  %%CITATION = PHRVA,D76,115016;%%


%\cite{Thaler:2008ju}
\bibitem{topjet}
  J.~Thaler and L.~T.~Wang,
  %``Strategies to Identify Boosted Tops,''
  JHEP {\bf 0807}, 092 (2008)
  [arXiv:0806.0023 [hep-ph]];
  %%CITATION = JHEPA,0807,092;%%
%\cite{Kaplan:2008ie}
%\bibitem{Kaplan:2008ie}
  D.~E.~Kaplan, K.~Rehermann, M.~D.~Schwartz and B.~Tweedie,
  %``Top Tagging: A Method for Identifying Boosted Hadronically Decaying Top
  %Quarks,''
  Phys.\ Rev.\ Lett.\  {\bf 101}, 142001 (2008)
  [arXiv:0806.0848 [hep-ph]];
  %%CITATION = PRLTA,101,142001;%
%\cite{Almeida:2008tp}
%\bibitem{Almeida:2008tp}
  L.~G.~Almeida, S.~J.~Lee, G.~Perez, I.~Sung and J.~Virzi,
  %``Top Jets at the LHC,''
  Phys.\ Rev.\  D {\bf 79}, 074012 (2009)
  [arXiv:0810.0934 [hep-ph]].
  %%CITATION = PHRVA,D79,074012;%%

%\cite{Pukhov:1999gg}
%\bibitem{Pukhov:1999gg}
\bibitem{CalcHEP}
  A.~Pukhov {\it et al.}, Preprint INP MSU 98-41/542;
  A.~Pukhov {\it et al.},
  %``CompHEP: A package for evaluation of Feynman diagrams and integration  over
  %multi-particle phase space. User's manual for version 33,''
  arXiv:hep-ph/9908288;
  %%CITATION = HEP-PH/9908288;%%
%\cite{Pukhov:2004ca}
%\bibitem{Pukhov:2004ca}
  A.~Pukhov,
  %``CalcHEP 3.2: MSSM, structure functions, event generation, batchs, and
  %generation of matrix elements for other packages,''
  arXiv:hep-ph/0412191.
  %%CITATION = HEP-PH/0412191;%%

%\cite{Maldacena:1997re}
\bibitem{Maldacena:1997re}
  J.~M.~Maldacena,
  %``The large N limit of superconformal field theories and supergravity,''
  Adv.\ Theor.\ Math.\ Phys.\  {\bf 2}, 231 (1998)
  [Int.\ J.\ Theor.\ Phys.\  {\bf 38}, 1113 (1999)]
  [arXiv:hep-th/9711200].
  %%CITATION = IJTPB,38,1113;%%

%\cite{ArkaniHamed:2000ds}
\bibitem{ArkaniHamed:2000ds}
  N.~Arkani-Hamed, M.~Porrati and L.~Randall,
  %``Holography and phenomenology,''
  JHEP {\bf 0108}, 017 (2001)
  [arXiv:hep-th/0012148].
  %%CITATION = JHEPA,0108,017;%%

%\cite{Rattazzi:2000hs}
\bibitem{Rattazzi:2000hs}
  R.~Rattazzi and A.~Zaffaroni,
  %``Comments on the holographic picture of the Randall-Sundrum model,''
  JHEP {\bf 0104}, 021 (2001)
  [arXiv:hep-th/0012248].
  %%CITATION = JHEPA,0104,021;%%





\end{thebibliography}
\end{document}